\documentclass[twocolumn,twocolappendix,usenames,dvipsnames]{aastex63}

\usepackage{amsmath}

\shorttitle{Entropy-conserving scheme for nonthermal energies}
\shortauthors{Semenov, Kravtsov, Diemer}

\graphicspath{{figures/}}

\usepackage{etoolbox}
\makeatletter
\patchcmd{\NAT@citex}
  {\@citea\NAT@hyper@{\NAT@nmfmt{\NAT@nm}\NAT@date}}
  {\@citea\NAT@nmfmt{\NAT@nm}\NAT@hyper@{\NAT@date}}
  {}
  {}
\patchcmd{\NAT@citex}
  {\@citea\NAT@hyper@{%
     \NAT@nmfmt{\NAT@nm}%
     \hyper@natlinkbreak{\NAT@aysep\NAT@spacechar}{\@citeb\@extra@b@citeb}%
     \NAT@date}}
  {\@citea\NAT@nmfmt{\NAT@nm}%
   \NAT@aysep\NAT@spacechar%
   \NAT@hyper@{\NAT@date}}
  {}
  {}
\patchcmd{\NAT@citex}
  {\@citea\NAT@hyper@{%
     \NAT@nmfmt{\NAT@nm}%
     \hyper@natlinkbreak{\NAT@spacechar\NAT@@open\if*#1*\else#1\NAT@spacechar\fi}%
       {\@citeb\@extra@b@citeb}%
     \NAT@date}}
  {\@citea\NAT@nmfmt{\NAT@nm}%
   \NAT@spacechar\NAT@@open\if*#1*\else#1\NAT@spacechar\fi%
   \NAT@hyper@{\NAT@date}}
  {}
  {}
\makeatother

\newcommand{\vect}[1]{{\pmb #1}}
\newcommand{\dotp}{\boldsymbol{\cdot}}
\newcommand{\grad}{\boldsymbol{\nabla}}

\newcommand{\divg}{\grad \dotp}

\def\Pth{P_{\rm th}}
\def\eth{e_{\rm th}}
\def\Sth{S_{\rm th}}
\def\gth{{\gamma_{\rm th}}}

\def\Pext{P_{\rm nt}}
\def\eext{e_{\rm nt}}
\def\Sext{S_{\rm nt}}
\def\gext{{\gamma_{\rm nt}}}

\def\etot{e_{\rm tot}}
\def\ekin{e_{\rm kin}}
\def\ecr{e_{\rm cr}}
\def\eturb{e_{\rm turb}}

\def\rs{r_{\rm s}}

\def\SSFR{{\dot{\Sigma}}_\star}
\def\rhoSFR{{\dot{\rho}}_\star}
\def\epsff{\epsilon_{\rm ff}}
\def\avir{\alpha_{\rm vir}}
\def\avirsf{\alpha_{\rm vir,sf}}
\def\cs{c_{\rm s}}
\def\sturb{\sigma_{\rm turb}}
\def\stot{\sigma_{\rm tot}}
\def\nsf{n_{\rm sf}}

\def\SFR{\dot{M}_\star}
\def\Mg{{M}_{\rm g}}
\def\Msf{{M}_{\rm sf}}
\def\tff{t_{\rm ff}}
\def\fsf{f_{\rm sf}}

\def\Nc{N_{\rm c}}

\def\tnsf{t_{\rm nsf}}
\def\tglob{\tau_{\rm dep}}

\def\Lstar{$L_\star$}
\def\ns{$n$--$\stot$}

\def\pc{{\rm\;pc}}
\def\kpc{{\rm\;kpc}}
\def\Mpc{{\rm\;Mpc}}
\def\Myr{{\rm\;Myr}}

\def\Msun{{\rm\;M_\odot}}

\def\Msunpc2{{\rm\;M_\odot\;pc^{-2}}}
\def\kms{{\rm\;km\;s^{-1}}}
\def\cc{{\rm\;cm^{-3}}}
\def\cm2s{{\rm\;cm^{2}\;s^{-1}}}

\def\K{{\rm\;K}}

\begin{document}

\title{Entropy-Conserving Scheme for Modeling Nonthermal Energies in Fluid Dynamics Simulations}

\author[0000-0002-6648-7136]{Vadim A. Semenov}
\altaffiliation{\href{mailto:vadim.semenov@cfa.harvard.edu}{vadim.semenov@cfa.harvard.edu} \\ NHFP Hubble Fellow.}
\affiliation{Center for Astrophysics $|$ Harvard \& Smithsonian, 60 Garden St, Cambridge, MA 02138, USA}

\author[0000-0003-4307-634X]{Andrey V. Kravtsov}
\affiliation{Department of Astronomy \& Astrophysics, The University of Chicago, Chicago, IL 60637, USA}
\affiliation{Kavli Institute for Cosmological Physics, The University of Chicago, Chicago, IL 60637, USA}
\affiliation{Enrico Fermi Institute, The University of Chicago, Chicago, IL 60637, USA}

\author[0000-0001-9568-7287]{Benedikt Diemer}
\affiliation{Department of Astronomy, University of Maryland, College Park, MD 20742, USA}

\begin{abstract}
We compare the performance of energy-based and entropy-conserving schemes for modeling nonthermal energy components, such as unresolved turbulence and cosmic rays, using idealized fluid dynamics tests and isolated galaxy simulations. While both methods are aimed to model advection and adiabatic compression or expansion of different energy components, the energy-based scheme numerically solves the nonconservative equation for the energy density evolution, while the entropy-conserving scheme uses a conservative equation for modified entropy. Using the standard shock tube and Zel'dovich pancake tests, we show that the energy-based scheme results in a spurious generation of nonthermal energy on shocks, while the entropy-conserving method evolves the energy adiabatically to machine precision. We also show that, in simulations of an isolated \Lstar~galaxy, switching between the schemes results in $\approx 20\text{--}30\%$ changes of the total star formation rate and a significant difference in morphology, particularly near the galaxy center. We also outline and test a simple method that can be used in conjunction with the entropy-conserving scheme to model the injection of nonthermal energies on shocks. Finally, we discuss how the entropy-conserving scheme can be used to capture the kinetic energy dissipated by numerical viscosity into the subgrid turbulent energy {\it implicitly}, without explicit source terms that require calibration and can be rather uncertain. Our results indicate that the entropy-conserving scheme is the preferred choice for modeling nonthermal energy components, a conclusion that is equally relevant for Eulerian and moving-mesh fluid dynamics codes.
\end{abstract}


\section{Introduction}

Stellar feedback plays a critical role in shaping properties of simulated galaxies on all relevant scales, from the overall stellar mass and star formation rate to metallicity, galaxy morphology, and chemical abundances in the circumgalactic (CGM) and intergalactic medium \citep[IGM; e.g.,][see also \citealt{Naab.Ostriker.2017} and \citealt{Vogelsberger.etal.2020} for recent reviews]{Governato.etal.2010,Brook.etal.2012,Agertz.etal.2013,Hopkins.etal.2013,Hopkins.etal.2014,Stinson.etal.2013,Agertz.Kravtsov.2015,Agertz.Kravtsov.2016}. Modeling feedback processes in numerical simulations of galaxy formation has been the focus of extensive theoretical efforts. 

Over the past decades a number of different methods have been developed to model stellar feedback, ranging from the direct injection of energy and momentum to the interstellar medium (ISM) when the physical scales of this injection can be resolved \citep[e.g.,][]{Hopkins.etal.2011,Hopkins.etal.2012,Hopkins.etal.2017b,Agertz.etal.2013,Marinacci.etal.2019,Jeffreson21-HIIfb,Smith.etal.2020}, to effective models of the ISM and outflow driving when the multiphase ISM structure cannot be sufficiently resolved \citep[e.g.,][]{Yepes.etal.1997,Springel.Hernquist.2003,Braun.Schmidt.2012,Vogelsberger.etal.2013,Springel18}.

One important class of feedback implementations is based on explicit modeling of a nonthermal energy component sourced by young stars in addition to gas thermal energy \citep[e.g.,][]{Springel.2000,Agertz.etal.2013,Teyssier.etal.2013,Benincasa.etal.2016}. Common examples of such nonthermal components are cosmic rays \citep[e.g.,][]{Jubelgas08,Uhlig.etal.2012,Booth.etal.2013,Salem.Bryan.2014,Salem.etal.2014,Pakmor.etal.2016,Ruszkowski.etal.2017,Wiener.etal.2017,Chan.etal.2019,Buck.etal.2020} and small-scale ISM turbulence \citep[e.g.,][]{Braun.etal.2014,Schmidt.etal.2014,Semenov.etal.2016,Semenov.etal.2021,Kretschmer2020,Kretschmer2021}. 
Such models differ by their sink, source, and extra transport terms (such as cosmic ray transport or turbulent diffusion). Without these terms, the behavior of the nonthermal energy is analogous to the thermal energy and consists of advection with gas density and the ``$PdV$'' work done during gas compression or expansion. 

Implementation-wise, nonthermal energies can be modeled using the methods developed for modeling the thermal energy of the gas. 
Such methods were introduced to follow gas temperature in highly supersonic flows typical for galaxy formation simulations\footnote{For example, the motion of the cold ISM due to galactic rotation and translational velocity ($\cs\sim 1\text{--}3 \kms$, $v \sim 100\text{--}300 \kms$), accretion of cold gaseous streams at high redshifts, and other motions of $T \sim 10^4\K$ gas in the IGM ($\cs\sim10\kms$, $v \sim $~few 100$\kms$) are flows with extremely high Mach numbers, $v/\cs > 30$.} where the modeling of thermal energy requires special attention. Indeed, when the flow is subsonic or only moderately supersonic, the thermal energy can be derived as the difference between the total and kinetic energy, $\eth = \etot-\ekin$, which are readily available in conservative methods that follow the gas mass, momentum, and total energy. However, when the flow is highly supersonic, the difference $\etot-\ekin$ becomes small and comparable to the truncation errors of $\etot$ and $\ekin$, requiring a more accurate method for modeling $\eth$.

Two such methods for modeling thermal energy have been introduced in the early days of hydrodynamic galaxy formation simulations. In the ``dual-energy’’ formalism proposed by \citet{Bryan.etal.1995}, thermal energy is modeled as a separate variable in addition to total energy, with the $PdV$ source term being computed explicitly. The second method, proposed by \citet{Ryu.etal.1993}, models the gas \emph{entropy} as a separate variable, which then can be used to compute temperature or thermal energy.  The key advantage of the latter method is that gas entropy obeys a conservative equation, at least outside of shocks and in regions of the flow where dissipative processes are negligible. The entropy equation can thus be solved accurately without  explicit source terms that can introduce significant errors.

Both these methods involve several parameters including the criteria that are used to decide whether $\eth$ should be taken from the explicitly modeled variable or synchronized with the total energy as $\eth = \etot-\ekin$ \citep{Ryu.etal.1993,Bryan.etal.1995,Springel.2010,Teyssier.2015}. In most idealized problems, both methods and different synchronization criteria work comparably well, and their differences tend do be small even in more realistic problems \citep[e.g.,][]{Costa.etal.2020}. However, in some cases, these choices can have significant effects. For example, \citet{Villasenor.etal.2020} showed that the formation of shock-heated gaseous halos in the IGM can be strongly delayed by the choice of the criteria used to synchronize thermal and total energies.

The choice of the method for modeling of \emph{nonthermal} components may have an even more significant impact. Indeed, the special treatment of thermal energy is required only in the extreme regime of highly supersonic flows, which are typically limited to a small fraction of the problem volume and where thermal pressure is by definition negligible. In contrast, nonthermal energies must be followed separately from the total energy in the entire simulation domain. Therefore, the choice of the method can impact results via the dynamical effect of nonthermal pressure and via its effect on other physical processes---such as star formation, radiative cooling, chemistry, etc.---which are sometimes coupled to the nonthermal energy.

In this paper, we explore the impact of different choices in modeling nonthermal energy on the evolution of gas using a series of idealized tests and more realistic simulations of an isolated \Lstar~galaxy. In particular, we compare the two methods in shock tube and hydrodynamic Zel'dovich pancake tests and show that the energy-based method results in the spurious generation of nonthermal energy on shocks while the entropy-conserving method can enforce the intended adiabatic behavior. A similar behavior was found by \citet{Kudoh.Hanawa.2016} and \citet{Gupta.etal.2021} in the context of numerical modeling cosmic rays as a nonthermal energy component. Here we further show that the entropy errors increase significantly for nonthermal components with larger adiabatic indices (such as subgrid turbulence with $\gamma=5/3$ compared to cosmic rays with $\gamma=4/3$), and for strongly compressive shocks. The latter becomes particularly problematic for modeling nonthermal components in galaxy formation simulations where highly compressive shocks are ubiquitous due to efficient radiative cooling in the dense ISM.

The paper is organized as follows. In Section~\ref{sec:schemes}, we review the methods of \citet{Bryan.etal.1995} and \citet{Ryu.etal.1993} and provide additional details about our implementation of these methods in the ART galaxy formation code. We also outline a simple method that can be used to implement the injection of nonthermal energy on shocks (Section \ref{sec:schemes:injection}). In Section~\ref{sec:tests}, we present the results of shock tube and hydrodynamic Zel'dovich pancake tests. Then, in Section~\ref{sec:galaxy}, we further compare the methods in simulations of an isolated \Lstar~galaxy and show that the choice of the method can significantly impact galaxy properties. In Section~\ref{sec:discussion}, we discuss our results and propose a novel method for modeling unresolved turbulent energy by capturing numerical dissipation with the entropy-conserving scheme. We summarize our conclusions in Section~\ref{sec:summary}.

\section{Implementation of thermal and nonthermal energies}
\label{sec:schemes}

To explore the impact of specific choices made in modeling nonthermal energy components, we use idealized tests and more realistic simulations of an isolated galaxy performed with the adaptive mesh refinement (AMR) hydrodynamics and $N$-body code ART \citep{Kravtsov.1999,Kravtsov.etal.2002,Rudd.etal.2008,Gnedin.Kravtsov.2011}. 
The hydrodynamic fluxes in the ART code are handled by a second-order Godunov-type method \citep{Colella.Glaz.1985} with a piecewise linear reconstruction of states at the cell interfaces \citep{vanLeer.1979} and a monotonized central slope limiter based on \citet{Colella.1985}.

Here we summarize our implementation of different methods for solving the set of advection equations with $PdV$ work source terms for the thermal and nonthermal energy components. The key assumption is that the adiabatic evolution of these components can be modeled separately from any nonadiabatic processes, which can be added via source terms. Thus, during a hydrodynamics step and before the source terms are applied, the internal energies of the thermal and nonthermal components evolve adiabatically:  
\begin{align}
\label{eq:eth} \frac{\partial \eth}{\partial t} + \divg (\eth \vect{u}) &= -\Pth \divg \vect{u}, \\
\label{eq:eext} \frac{\partial \eext}{\partial t} + \divg (\eext \vect{u}) &= -\Pext \divg \vect{u}.
\end{align}
Here and throughout the paper, $\eth$ and $\eext$ denote energies per unit volume, and $\grad \equiv (\partial/\partial x, \partial/\partial y, \partial/\partial z)$ is the gradient operator.
For clarity, we consider a single nonthermal energy component, $\eext$, but all of the results and conclusions can be trivially generalized for an arbitrary number of such energy components.\footnote{For multiple nonthermal components, $e_{{\rm nt,}i}$, each component $i$ obeys Equation~(\ref{eq:eext}) and its individual equation of state ($P_{{\rm nt,}i} = (\gamma_{{\rm nt,}i} - 1) e_{{\rm nt,}i}$, in our implementation); the total pressure of gas becomes $P = \Pth + \sum_i P_{{\rm nt,}i}$, and $\eext$ in the expression for $\etot$ and Equation~(\ref{eq:eth-sync}) is replaced by $\sum_i e_{{\rm nt,}i}$.}
In our calculations we assume an ideal gas equation of state for each of the components: $P_i = (\gamma_i-1)e_i$ with $i\in [{\rm th},{\rm nt}]$.

The above set of equations is coupled with gas dynamics via the total pressure $P=\Pth + \Pext$. In addition, as thermal and nonthermal energies contribute to the gas total energy, $\etot$, a suitable method should be adopted to ensure that $\etot = \ekin + \eth + \eext$, where $\ekin = \rho\vect{u}^2/2$, when such synchronization is appropriate.

The choice of the synchronization method is dictated by the expected behavior of nonthermal energies across shocks. Indeed, in shocked regions, the difference $\etot - \ekin$ contains the adiabatic change in $\eth$ and $\eext$ and the kinetic energy converted into thermal energy by the shock. In the absence of a subgrid model for nonthermal energy generation within shocks, the conservative choice is to assume that all nonadiabatic energy increase across shocks is thermalized,
\begin{equation}
\label{eq:eth-sync}
    \eth = \etot-\frac{\rho \vect{u}^2}{2} - \eext,
\end{equation}
while the energies of the nonthermal components change adiabatically.

In this work, we consider the assumption that nonthermal components change adiabatically as the most ``basic,’’ conservative choice as this is the behavior implied by the underlying equations ((\ref{eq:eth}) and (\ref{eq:eext})) for smoothed flows. Any other, nonadiabatic behavior can be taken into account in the partitioning of $\etot - \ekin$ between $\eth$ and $\eext$ (Section~\ref{sec:schemes:injection}).
In particular, nonthermal energies can be generated within shocks, e.g., via cosmic ray acceleration and turbulence driving. It is also worth noting that the shock structure and jump conditions are generally expected to be modified by kinetic non-MHD effects and propagation of cosmic rays across the shock \citep[e.g.,][]{Voelk.etal.1984,Bret2020,Haggerty.Caprioli.2020,Tsung.etal.2020}. These modifications are uncertain and depend on the details and physical parameters of shocks, which generally are not possible to model self-consistently in cosmological simulations. Thus, the effects of such modifications need to be modeled phenomenologically as a part of the CR shock injection model. This was referred to as the closure problem of the shock solution in the presence of nonthermal energy components in a recent study by \citet{Gupta.etal.2021}.  

In the remainder of this section, we outline two methods implemented in the ART code for solving Equations~(\ref{eq:eth})--(\ref{eq:eext}) and our choices of synchronization criteria that determine when $\eth$ is reset using Equation~(\ref{eq:eth-sync}).
These methods were originally developed for modeling thermal energy in highly supersonic flows by \citet{Bryan.etal.1995} and \citet{Ryu.etal.1993}, and in this paper we explore their behavior in the context of modeling nonthermal energies. In principle, different energy components can be treated using different methods, however, we opt for a consistent treatment of all components using the same method.

\subsection{Energy-based scheme}
\label{sec:schemes:energy}

In the ``dual-energy'' formulation proposed by \citet{Bryan.etal.1995}, Equations~(\ref{eq:eth})--(\ref{eq:eext}) are solved in their original form: the energies $\eth$ and $\eext$ are advected using conservative mass fluxes, and the $PdV$ work is added as a source term, making the scheme nonconservative. 

In the ART code, we advect $\eth$ and $\eext$ as passive scalars: we interpolate specific energies, $e_i/\rho$ with $i\in [{\rm th},{\rm nt}]$, at the upwind side of the interface using the same reconstruction scheme as for all other hydrodynamic variables. The advection flux is then obtained by multiplying these reconstructed values by the average mass flux across that cell interface computed by the Riemann solver. 
This method ensures that the advection fluxes of $\eth$ and $\eext$ are consistent with the mass flux, which enables the code to advect contact discontinuities to machine precision (see Appendix~\ref{app:pressure-balance} and, in particular, the middle panel of Figure~\ref{fig:pressure-balance}).

In Godunov-type methods, the $PdV$ source terms can be computed relatively accurately and efficiently because the time-averaged velocity of gas at a given interface is computed during the solution of the Riemann problem across cell interfaces and can be used for an accurate estimate of $\divg \vect{u}$.
In the ART code, $\divg \vect{u}$ is computed by accumulating the fluxes of average velocities at cell interfaces and applying the Gauss--Ostrogradsky divergence theorem for each cell. Using this estimate of $\divg \vect{u}$, the source term is applied to the energy components after the advection step.

This method of advancing $\eth$ and $\eext$ is used in regions where the flow is highly supersonic and the values of $\eth$ from $\etot$ can become smaller than the truncation error of the scheme, meaning highly inaccurate. In regions of modest Mach numbers, $\eth$ is synchronized with $\etot$ using Equation~(\ref{eq:eth-sync}). Specifically, the energies are synchronized in cells with

\begin{equation}
\label{eq:b95-sync1}
    \frac{\etot-\ekin}{\etot} > \eta_1,
\end{equation}
with the value of $\eta_1 = 10^{-3}$ suggested by \citet{Bryan.etal.1995}. This condition is equivalent to the Mach number threshold of $M < \sqrt{2(\eta_1^{-1}-1)/(\gamma(\gamma-1))} \approx 40$.\footnote{In the absence of nonthermal energy components, this relation follows from $\etot=\ekin+\eth$ and $\ekin = \rho u^2/2 = M^2 \rho \cs^2/2 = M^2 \gamma (\gamma - 1) \eth /2$. In the presence of nonthermal components, $M$ and $\gamma$ correspond to effective values.} 
Additionally, \citet{Bryan.etal.1995} apply a second criterion based on the Mach numbers of adjacent cells:

\begin{equation}
\label{eq:b95-sync2}
    \frac{e_{{\rm tot},j}-e_{{\rm kin},j}}{\max(e_{{\rm tot},j-1},e_{{\rm tot},j},e_{{\rm tot},j+1})} > \eta_2,
\end{equation}
where $j$ denotes the cell to which the criterion is applied and $j\pm1$ are its immediate neighbors. We use the value of $\eta_2 = 0.1$ as in the original scheme of  \citet{Bryan.etal.1995}, which corresponds to $M\lesssim4$ and makes this a significantly more conservative criterion than Equation~(\ref{eq:b95-sync1}).

Overall, these two criteria are used to select the method by which $\eth$ is evolved---i.e., either by updating its value from $\etot$ (Equation~(\ref{eq:eth-sync})), or by solving Equation~(\ref{eq:eth}) independently of $\etot$. We choose to always follow $\eext$ using Equation~(\ref{eq:eext}). For smaller $\eta_1$ and $\eta_2$, $\eth$ is evolved synchronously with $\etot$ in a larger fraction of the simulated volume, which in the limiting case of $\eta_1 = \eta_2 = 0$ corresponds to not using the dual-energy formulation at all.

Note that our implementation slightly differs from the original scheme of \citet{Bryan.etal.1995}, in which  the first criterion is only used to identify the cells in which thermal pressure is computed using explicitly advected $\eth$, while only the second criterion is used to decide whether to synchronize $\eth$ and $\etot$. This is done to limit the dynamical effect of explicitly advected $\eth$ on the solution. However, given that the first criterion selects only the gas with $\eth \ll \etot$, the modeling of $\eth$ in such gas should have a negligible effect on the dynamics of the gas. We thus opt to use the first criterion to synchronize $\eth$ with $\etot$ and always compute pressures from explicitly advected $\eth$ and $\eext$.

Finally, we also find that modeling of contact discontinuities in presence of multiple energy components with different adiabatic indices can be significantly improved if the total energy flux is computed consistently with the advection fluxes of $\eth$ and $\eext$ described above. The total flux of thermal+nonthermal energy produced by the Riemann solver at a given interface is $u P/(\gamma-1)$, where $\gamma = P/e+1$, with $P=\Pth+\Pext$ and $e=\eth+\eext$, is the adiabatic index of gas which is explicitly followed by our Riemann solver \citep{Colella.Glaz.1985}. Away from shocks, this flux corresponds to the total advection flux of $\eth+\eext$. However, the numerical value of this flux does not necessarily equal to the total flux computed by the advection scheme. We find that, even though this inconsistency is small, it still can produce significant artifacts in idealized tests like the shock tube test or advection of a pressure balance mode.

To make the total energy flux consistent with the advection scheme, we subtract $u P/(\gamma-1)$ from the Riemann solution for the flux of $e_{\rm tot}$ and add the total flux of $\eth+\eext$ computed by the advection scheme. This correction should be done only outside shocked regions where the Riemann solution for pressure does not contain kinetic energy dissipated by shocks. For this reason, we apply this correction only on the interfaces where these two estimates of the total advection flux are within 3\% of each other. By design, this correction results only in a small adjustment of the total energy flux, and the scheme remains explicitly energy conserving. We also find that this correction is necessary only in presence of multiple energy components with different adiabatic indices.

The necessity for the above correction is motivated by the design of the \citet{Colella.Glaz.1985} Riemann solver as it operates with the total pressure and effective adiabatic indices on the left and right sides of cell interfaces. Because of this, the total flux of thermal and nonthermal energy produced by the Riemann solver will generally be different from the same flux computed by the advection scheme. An alternative solution is to develop a Riemann solver that would operate on individually reconstructed pressure components and follow their evolution across the interface. To this end, one can use analytic solutions for the evolution of the nonthermal component with or without injection on the shock \citep[see Appendix C in][]{Pfrommer.etal.2017} or even try including kinetic non-MHD effects and propagation of cosmic rays \citep[e.g.,][]{Voelk.etal.1984,Haggerty.Caprioli.2020,Bret2020,Tsung.etal.2020}. Such a Riemann solver will provide the solution for fluxes of different energy components without requiring a separate advection scheme. Note, however, that to advance thermal and nonthermal energies one still will be required to compute the source terms, which, as we will show below, can generate significant errors in shocked regions.

\subsection{Entropy-conserving scheme}
\label{sec:schemes:entropy}

\citet{Ryu.etal.1993} proposed an alternative method, where a modified entropy, $\rho S_i \equiv P_i/\rho^{\gamma_i-1}$, is followed as a separate variable instead of energy. The equations for $\rho S_i$ can be derived by combining Equations~(\ref{eq:eth})--(\ref{eq:eext}) with the continuity equation,
\begin{equation}
\label{eq:rho} \frac{\partial \rho}{\partial t} + \divg (\rho \vect{u}) = 0,
\end{equation}
which gives
\begin{align}
\label{eq:Sth} \frac{\partial \rho\Sth}{\partial t} + \divg (\rho\Sth \vect{u}) &= 0, \\
\label{eq:Sext} \frac{\partial \rho\Sext}{\partial t} + \divg (\rho\Sext \vect{u}) &= 0.
\end{align}

These equations are in the conservative form and thus can be solved with a numerical scheme that conserves $\rho\Sth$ and $\rho\Sext$ to machine precision.\footnote{In fact, as was pointed out by \citet{Ryu.etal.1993}, one can construct an arbitrary number of conservative quantities of the form $P^\alpha \rho^{1-\gamma \alpha}$ with $\alpha \in (-\infty, \infty)$. This can be easily checked by expanding the conservation equation for such a quantity and using Equations~(\ref{eq:eth}), (\ref{eq:eext}), and (\ref{eq:rho}). The entropy conservation form corresponds to $\alpha=1$. Another special case of $\alpha=1/\gamma$, leading to a conserved variable $P^{1/\gamma}$, was explored by \citet{Kudoh.Hanawa.2016}.} Given that the gas density can also be conserved to machine precision, this means that we can get accurate estimates of the entropy of the thermal and nonthermal energy components, $\Sth$ and $\Sext$, using $\rho S$ and $\rho$ and computing the entropy fluxes, $S \rho \vect{u}$, consistently with the mass fluxes, $\rho \vect{u}$. 

Implementation-wise, we use the same advection scheme for entropy as for the energy-based scheme described above, with specific entropy being advected as a passive scalar.
Specifically, to compute the advection flux of entropy across a given cell interface, we interpolate the value of $S_i = P_i/\rho^{\gamma_i}$ (which can be thought of as a proxy for entropy per unit mass) at the upwind side of the interface using the same reconstruction scheme as for all other variables and multiply it by the average mass flux from the solution of the Riemann problem \citep[][]{Springel.2010}.\footnote{We also tried using the ``raw'' value of $S_i$ from the upwind cell without reconstruction but found that using reconstructed values of $S_i$ better preserves sharp contact discontinuities in different pressure components.}

The criteria for synchronizing $S_i$ with the total energy of the gas, $\etot$, also differ qualitatively from those in the energy-based method. Indeed, thermal entropy can be generated in shocks, and Equation~(\ref{eq:Sth}) therefore becomes invalid in regions with strong shocks, meaning that $\etot$ must be used instead. Thus, apart from the threshold on the Mach number of the flow, additional criteria must be used to identify shocked regions. For example, \citet{Ryu.etal.1993} check if the flow is locally converging ($\divg \vect{u} < 0$) and the pressure jump across the cell is larger than a chosen threshold value.

An alternative method to identify shocked regions was proposed by \citet[][Section~3.5]{Springel.2010}. This method is based on the idea that in Godunov-type methods the production of entropy is taken care of by the shock(s) present in the Riemann solution for the hydrodynamic fluxes across cell interfaces. For weak shocks, the dissipated energy is strongly subdominant to the adiabatic increase in the internal energy, and therefore one can identify the regions with strong production of entropy by defining a threshold in the Mach number of such ``Riemann shocks.'' \citet{Springel.2010} suggests the value $M_{\rm R,crit}=1.1$, so that $\etot$ is used in the regions where $M_{\rm R} > M_{\rm R,crit}$ on at least one of the cell's interfaces. Although the relation between the shocks in the Riemann solutions and the physical shocks in the simulated flow is not direct, the total entropy generated by the real shock accumulates from the increments produced by the ``Riemann shocks'' on the interfaces resolving the real shock, which warrants the above criterion.

We determine the value of $M_{\rm R}$ at a given interface from the total pressure jumps present in the solution of the Riemann problem at this interface. Such a solution consists of the left (l) and right (r) states sandwiching the so-called ``star’’ region ($\star$) and separated by either shock(s) and/or expansion fans \citep[e.g.,][]{toro}. The relevant pressure jump, $j_P$, is defined as the maximum of $P_\star/P_{\rm l}$ and $P_\star/P_{\rm r}$. The corresponding Mach number can then be calculated from the Rankine--Hugoniot jump conditions, with pre- and post-shock regions corresponding to the cell on the side with the largest pressure jump (denoted with a subscript ``$j$'') and the ``star'' region, respectively \citep[see also][]{Pfrommer.etal.2017}:
\begin{align}
    M^2_{\rm R} &= \frac{(j_P-1)}{\gamma_{\rm eff}}\frac{A}{A-B}, \label{eq:MR} \\
    A &\equiv (\gamma_j-1)[(\gamma_\star+1)j_P+(\gamma_\star-1)], \nonumber \\
    B &\equiv (\gamma_\star-1)[(\gamma_j-1)j_P + (\gamma_j+1)], \nonumber
\end{align}
where $\gamma_{\rm eff} \equiv (\gth P_{{\rm th},j}+\gext P_{{\rm nt},j})/(P_{{\rm th},j}+P_{{\rm nt},j})$ is the effective adiabatic index in the cell on the side with the largest pressure jump, and $\gamma_i \equiv P_i/e_i+1$ with $i \in [j,\star]$, and $P$ and $e$ being the total pressure and thermal+nonthermal energy in the corresponding regions. In our implementation, $\gamma_j$ is computed using the values of $P$ and $e$ from the corresponding cell, while the value of $\gamma_\star$ is taken directly from the solution of the \citet{Colella.Glaz.1985} Riemann solver, which solves for the variation in $\gamma$ explicitly. In the case when $\gamma_j = \gamma_\star \equiv \gamma$, Equation~(\ref{eq:MR}) can be simplified as
\begin{equation}
    M^2_{\rm R} = \frac{(\gamma+1)j_P+(\gamma-1)}{2\gamma}.
\end{equation}
We use this simplified form when $2|\gamma_j - \gamma_\star|/(\gamma_j + \gamma_\star)<0.01$, following  \citet{Pfrommer.etal.2017}. Having an estimate of the Mach number of shocks present in the Riemann solution at all cell interfaces, we define $M_{\rm R}$ for a cell as the maximum $M_{\rm R}$ value on that cell's interfaces.

In our experiments, we find that the $M_{\rm R,crit}$ threshold alone cannot prevent spurious heating in regions with strong velocity gradients. Such gradients can be interpreted by the Riemann solver as a discontinuity producing a sufficiently strong shock with a Mach number that can satisfy the above criterion. To filter out such cases, we require that the velocity of the shock(s) in the Riemann solution must be sufficiently large compared to the local flow velocity. In addition, to avoid injection of entropy in poorly resolved highly supersonic flows, we also require that the right-hand side of Equation~(\ref{eq:eth-sync}) is larger than a certain fraction of $\etot$, which is analogous to the criteria used by \citet{Ryu.etal.1993} and \citet{Bryan.etal.1995}.
Note that the latter criterion may not be required in moving-mesh codes \citep[e.g.,][]{Springel.2010} and when hydrodynamic equations are solved in a locally comoving frame \citep[e.g.,][]{Trac.Pen.2004}, where a significant fraction of kinetic energy is absorbed by the bulk flow of gas. For static-mesh Eulerian codes, however, this additional criterion is required to avoid artifacts in strongly supersonic flows.

Based on extensive experiments we find that the following set of criteria works well to identify shock regions: 
\begin{align}
\label{eq:crit-Mach} \max(M_{\rm R}) &> 1.1, \\
\label{eq:crit-vshock} |s_{\rm R,max}|/|v_{\rm cell}| &> 0.1, \\
\label{eq:crit-ediff} (\etot-\ekin-\eext)/\etot &> 10^{-3},
\end{align}
where $\max(M_{\rm R})$ is the maximal Mach number of the ``Riemann shocks'' at the interfaces of a given cell, $|s_{\rm R,max}|$ is the corresponding shock velocity, and $|v_{\rm cell}|$ is the gas velocity in the cell. When all of the above criteria are satisfied, the thermal energy is synchronized with total energy according to Equation~(\ref{eq:eth-sync}), meaning that thermal entropy is injected into the cell in the amount corresponding to the energy dissipated by the shock. Otherwise, the evolution of thermal energy is followed using the entropy conservation Equation~(\ref{eq:Sth}). To enforce the adiabatic behavior of nonthermal energy components, their evolution is always followed by Equation~(\ref{eq:Sext}).

As was pointed out by \citet{Springel.2010}, one significant disadvantage of such a scheme is that it forfeits strict conservation of energy. In structure and galaxy formation simulations, however, the total energy is usually not conserved anyway owing to radiative cooling, radiative or feedback heating, star formation, etc. All of these processes are uncertain and modeled approximately. At the same time, as we show below, in the idealized tests, the entropy-conserving scheme performs either comparably or better than the energy-based scheme.

If the strict conservation of energy is nevertheless desirable, one can disregard the solution for the thermal component and compute $\eth$ from $\etot$ while also following entropies of nonthermal components \citep{Kudoh.Hanawa.2016,Gupta.etal.2021}. Note that solving the equation for thermal entropy simultaneously with $\etot$ is still desirable, so that gas temperature and pressure can be evaluated in regions with highly supersonic flows. In practice, such a scheme can be achieved by relaxing criteria~(\ref{eq:crit-Mach}) and (\ref{eq:crit-vshock}) and using only Equation~(\ref{eq:crit-ediff}), so that explicitly modeled thermal entropy is used in highly supersonic regions. Such a scheme will ensure the conservation of both total energy and nonthermal entropy.

\subsection{Generation of nonthermal energies in shocks}
\label{sec:schemes:injection}

Synchronizing $\eth$ with $\etot$ according to Equation~(\ref{eq:eth-sync}) implies that all kinetic energy dissipated by shocks is thermalized. However, the synchronization procedure can easily be modified to account for nonthermal energy generation on shocks, such as cosmic ray acceleration or driving of small-scale turbulence when these components are modeled as $\eext$. Indeed, if the adiabatic evolution of $\eth$ and $\eext$ can be enforced during a hydrodynamic step, then at the end of the step the difference $e_{\rm diss} \equiv \etot-\ekin-\eth-\eext$ in each cell will correspond to the total energy dissipated by shocks during the step. Therefore, to convert a fraction $\zeta$ of this energy into a nonthermal component, one only needs to add $\zeta e_{\rm diss}$ to $\eext$ and the remaining $(1-\zeta) e_{\rm diss}$ to $\eth$. For $\zeta = 0$, this scheme is equivalent to using Equation~(\ref{eq:eth-sync}).

As we demonstrate in Section~\ref{sec:tests:injection}, this scheme works remarkably well when the adiabatic index of the injected energy is the same as that of the thermal energy, and it produces reasonable results when the adiabatic indices are different (e.g., for cosmic rays with $\gext=4/3$). At the same time, this scheme is trivial to implement within the entropy-conserving scheme as it requires only a minor modification of the energy synchronization algorithm, while a number of previous implementations of energy injection on shocks required on-the-fly shock-finding algorithms. Note that if $\zeta$ depends on the properties of the shock, as is generally the case for cosmic rays, the shock-finding algorithm is still needed to measure these properties for physical shocks. For example, one may try to estimate the local properties of the shock by considering several adjacent cells or using more complex algorithms \citep[e.g.,][]{Ryu.etal.2003,Pfrommer.etal.2006,Skillman.etal.2008,Schaal.Springel.2015}. However, the energy partitioning scheme can still be used to inject the corresponding nonthermal energy. 

This algorithm is qualitatively similar to the method outlined by \citet{Gupta.etal.2021}, where the total pressure is partitioned between the thermal and nonthermal components to maintain prescribed fractions in the regions identified as shocks. The main difference of the method described above is that only the kinetic energy locally dissipated by shocks is partitioned, while adiabatic compression is modeled separately by the entropy-conserving scheme. This can be advantageous for modeling weak shocks, in which adiabatic compression of the pre-shock thermal and nonthermal components contributes significantly to the post-shock pressure.
In principle, adiabatic compression can also be accounted for in the partitioning of the total pressure, however, the fractions of different pressure components depend on the shock Mach number and therefore it requires a shock-finding algorithm, while with the entropy-conserving method, the adiabatic behavior is built into the scheme.

\section{Idealized tests}
\label{sec:tests}

\subsection{Shock tube test with adiabatic nonthermal energy}
\label{sec:tests:shocktube}

\begin{figure}
\includegraphics[width=\columnwidth]{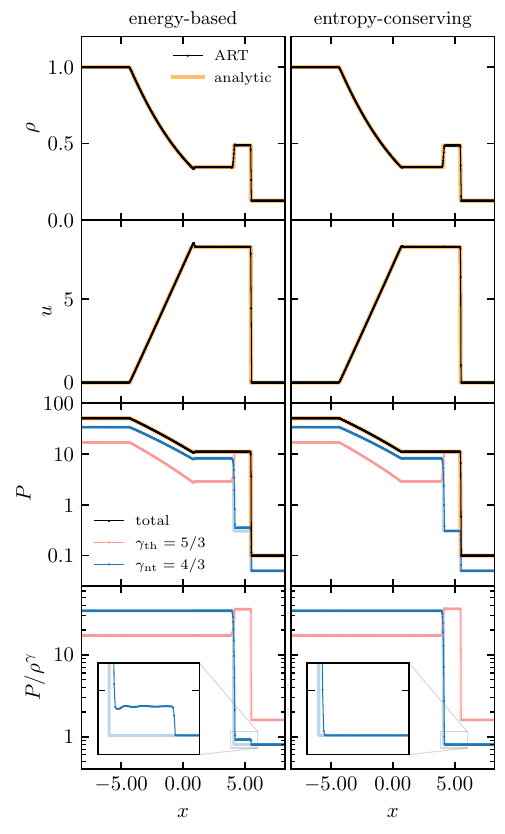}
\caption{\label{fig:shocktube} Shock tube test with an extra nonthermal energy representing cosmic rays ($\gext=4/3$) using the energy-based (left column) and entropy-conserving (right column) schemes for modeling thermal energy and cosmic rays. The initial conditions are taken from \citet{Pfrommer.etal.2017}: $(\rho,u,\Pth,\Pext) = (1,0,17.172,34.344)$ and $(0.125,0,0.05,0.05)$ for the left and right initial states, respectively. Thick lines show the analytic solutions from \citet[][Appendix C2]{Pfrommer.etal.2017}, with orange lines showing the solution for the density, velocity, and total pressure, and red and blue lines showing the solutions for the thermal and cosmic ray components, respectively. The results are compared at $t=0.5$. As the bottom panel shows, the energy-based scheme leads to a spurious production of nonthermal entropy at the shock, while the entropy-conserving scheme ensures nonthermal entropy conservation across the shock.}
\end{figure}

Figure \ref{fig:shocktube} shows the results of a shock tube test with an additional nonthermal fluid representing cosmic rays (CRs) with $\gamma_{\rm nt} = 4/3$ \citep{Pfrommer.etal.2006,Pfrommer.etal.2017}. The setup is analogous to the classic \citet{Sod.1978} test and has the same structure of the solution except that pressure consists of two components with different adiabatic indices, meaning that the effective $\gamma$ can vary from region to region. As the figure shows, both energy-based and entropy-conserving methods reproduce the analytic solution for the density, velocity, and total pressure, which demonstrates that the fluxes of the mass, momentum, and total energy are computed correctly in both schemes. 

\begin{figure}
\includegraphics[width=\columnwidth]{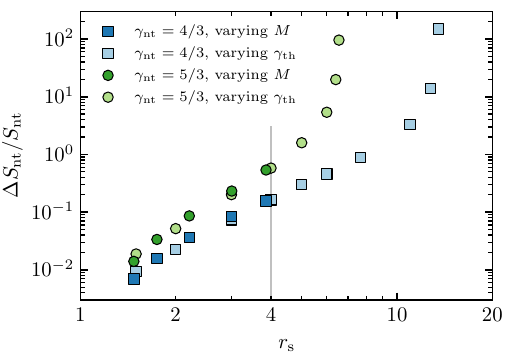}
\caption{\label{fig:S-error} Dependence of the nonthermal entropy error produced by the energy-based scheme in the post-shock region on the compression ratio of the shock, $\rs=\rho_1/\rho_0$, for different values of the adiabatic index of the nonthermal component: $\gext=4/3$ (blue squares) and $\gext=5/3$ (green circles). The entropy error increases monotonically with $\rs$ until $\Delta \Sext/\Sext$ becomes $\sim 1$, at which point the error diverges. For a larger $\gext$, this divergence occurs at a smaller $\rs$.
For this test, the nonthermal energy is dynamically decoupled from the hydrodynamics by making its initial value arbitrary small and removing its contribution from the total pressure.
The color intensity corresponds to different ways of varying $\rs$. Lighter colors correspond to variations in $\gth$, keeping the shock Mach number high so that $\rs \approx (\gth+1)/(\gth-1)$, with $\gth$ spanning a range from $\gth=5$ (leftmost points) to $\gth=1.16$ and $1.36$ (rightmost points for $\gext=4/3$ and $\gext=5/3$, respectively). 
Darker colors show the variation in the shock Mach number at fixed $\gth=5/3$ (from left to right, $M\approx1.3$, 1.5, 1.9, 3.0, 9.0), with the vertical gray line showing the high-Mach-number limit of $\rs=4$. }
\end{figure}

\begin{figure}
\centering
\includegraphics[width=1.967in]{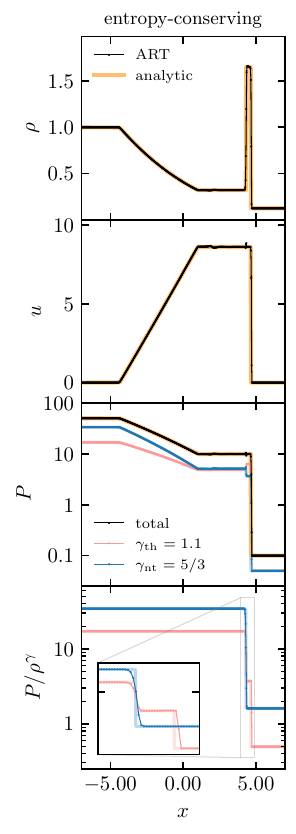}
\caption{\label{fig:shocktube-compressive} Results of the entropy-conserving scheme in the regime where the energy-based scheme is highly unstable. The initial conditions are the same as in Figure~\ref{fig:shocktube}, but the adiabatic indices of the thermal and nonthermal components are set to $\gth=1.1$ and $\gext=5/3$ to make the shock highly compressible, with a compression ratio of $\rs\approx13.2$. Such $\rs$ is significantly higher than the value at which the entropy error of the energy-based scheme diverges ($\rs \sim 6.5$ for $\gext=5/3$, see Figure~\ref{fig:S-error}). The entropy-conserving scheme still performs well. }
\end{figure}

The difference between the methods appears in the solution for the CR entropy and pressure as highlighted in the inset panels in the bottom row. In the energy-based scheme, CR entropy is spuriously produced at the shock, leading to a $\sim$20\% excess of the post-shock CR pressure. The post-shock thermal pressure becomes underestimated by the same (absolute) amount because in the post-shock region $\eth$ is computed from $\etot$ and $\ecr$ using Equation~(\ref{eq:eth-sync}). In contrast, the entropy-conserving scheme does not suffer from such errors: CR entropy is exactly conserved across the shock, resulting in correct CR and thermal pressures in the post-shock region.

This type of errors was also reported by \citet{Kudoh.Hanawa.2016} and \citet{Gupta.etal.2021}, who also showed that switching to a conservative equation for nonthermal components (cosmic rays in their studies) helps to avoid these errors. All these results indicate that the entropy errors are related to the treatment of the source term in the nonconservative equation of the energy-based scheme.

Indeed, the spurious production of nonthermal entropy in the energy-based scheme stems from the fact that inviscid Equations~(\ref{eq:eth}) and (\ref{eq:eext}) are formally invalid at shocks because of divergent gradients. In a numerical solution, however, shocks are smeared over several cells via numerical diffusivity and all gradients are finite. As the gas flows through such shocked regions, the $PdV$ source term computed from the local pressure and velocity gradients mixes the adiabatic compression with a fraction of the energy dissipated by the shock, leading to a nonadiabatic behavior.
In other words, solving Equations~(\ref{eq:eth}) and (\ref{eq:eext}) with explicit source terms results in injection of both thermal and nonthermal energy in the shocked regions in addition to the adiabatic increase due to gas compression.
In the absence of nonthermal energies, this effect does not introduce severe errors because, in the post-shock region, the solution for the advected $\eth$ is usually disregarded and $\eth$ is reset from the difference $\etot-\ekin$. In contrast, if nonthermal energies are present, only the sum $\eth+\eext$ can be set to $\etot-\ekin$ while the individual values of $\eth$ and $\eext$ become affected by the entropy error.

The $\sim 20\%$ error in the above test may seem small and unlikely to be important in real applications, e.g., galaxy formation simulations, due to much more uncertain source terms such as those involved in star formation and feedback modeling. However, we find that the error strongly increases for more compressible shocks and larger adiabatic indices of the nonthermal component. Indeed, as the entropy error originates from the $PdV$ source term ($\Pext \divg \vect{u} = \Pext \; \partial u/\partial x$ for a one-dimensional problem), its magnitude relative to $\eext$ increases with $\gext$ because $\Pext = (\gext-1) \eext$. The magnitude of the $PdV$ term also increases with the compression ratio of the shock ($\rs=\rho_1/\rho_0$, where ``0'' and ``1'' denote pre- and post-shock regions, respectively) because the magnitude of the velocity gradient increases with $\rs$, and the errors are amplified by stronger compression.

Figure~\ref{fig:S-error} shows the nonthermal entropy error in the post-shock region as a function of the compression ratio $\rs$ for $\gext=4/3$ (blue points) and $\gext=5/3$ (green points). To exclude the dynamical effect of the error on the solution we set the initial value of $\eext$ to an arbitrary small value ($<10^{-11}$ of $\eth$ in the post-shock region) and rerun the shock tube test with varying  shock Mach number (i.e., varying the initial pressure jump) and adiabatic indices $\gth$ and $\gext$.
Dark colors show the variation in the Mach number at fiducial $\gth=5/3$, which leads to variation in the compression ratio with the limiting value of $\rs = (\gth+1)/(\gth-1) = 4$ (shown with the vertical gray line). To explore variations of $\rs$ beyond this limiting value, we also rerun the tests with different $\gth$. Such a variation is relevant for practical applications, e.g., in galaxy formation simulations, because $\rs > 4$ can be achieved due to strong radiative cooling, even if $\gth$ is formally fixed at $5/3$.

The figure shows that, while the entropy errors of the $\gext=4/3$ component on strong shocks with $\rs \approx 4$ are $\sim 20\%$, the errors of the $\gext=5/3$ component increase by a factor of 4. Thus, the spurious production of nonthermal entropy becomes more problematic for modeling components such as small-scale ISM turbulence with $\gamma=5/3$ \citep[e.g.,][]{Robertson.Goldreich.2012,Schmidt.etal.2014}. Moreover, the error quickly increases as the shock becomes more compressive, and this increase saturates only when the post-shock $\eext$ becomes comparable to the total energy dissipated by the shock, independent of the pre-shock value of $\eext$. As the figure shows, for larger $\gext$, this divergence of errors occurs at smaller $\rs$: $\rs \sim 14$ and 6.5 for $\gext=4/3$ and 5/3, respectively. As we will show in Section~\ref{sec:galaxy}, such errors can significantly change the behavior of nonthermal components in galaxy formation simulations.

In the entropy-conserving scheme, the nonthermal entropy conservation equation is followed in the entire simulation domain. Figure~\ref{fig:shocktube-compressive} demonstrates that this scheme can produce an adequate solution even for strongly compressive shocks for which the energy-based scheme is highly unstable.

\subsection{Shock tube test with sourcing of nonthermal energy by the shock}
\label{sec:tests:injection}

\begin{figure}
\includegraphics[width=\columnwidth]{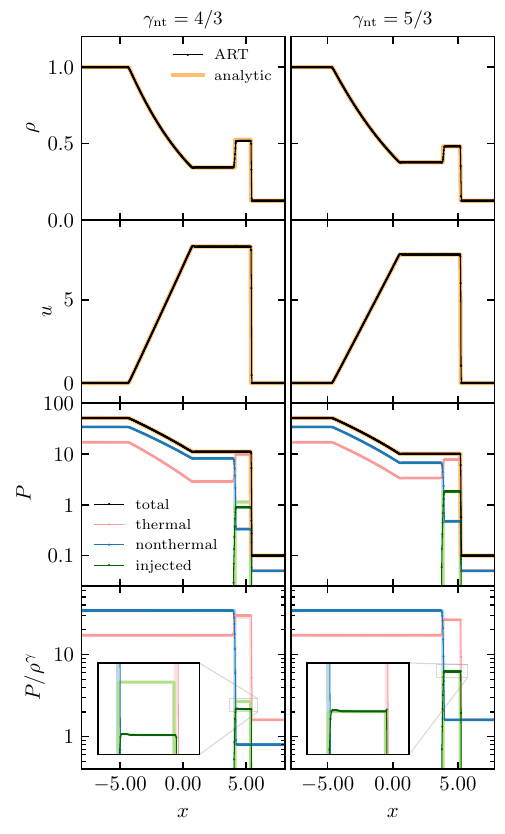}
\caption{\label{fig:shocktube-injection} Shock tube test with 20\% of dissipated energy injected in the nonthermal component using the entropy-conserving scheme and the energy partitioning method outlined in Section~\ref{sec:schemes:injection}. The initial conditions are the same as in Figure~\ref{fig:shocktube}. The left and right columns show the results for different adiabatic indices of the nonthermal component: $\gext=4/3$ and $5/3$, respectively. The nonthermal energy injected at the shock is tracked separately and shown with green color in the bottom two panels. The analytic solution for all quantities is computed following \citet[][Appendix C2]{Pfrommer.etal.2017} and is shown with the thick lines. When $\gext$ differs from $\gth=5/3$, the injected energy in the post-shock region is underestimated by $\sim 20\%$, while for $\gext=\gth=5/3$ the result is in perfect agreement with the analytical solution. }
\end{figure}

The shock tube test in the previous section assumes that all energy dissipated by the shock is thermalized, so that nonthermal energy behaves adiabatically. In this section, we switch to the case where nonthermal energy is also sourced by the shock as described in Section~\ref{sec:schemes:injection}. This method requires an accurate modeling of the adiabatic compression of both $\eth$ and $\eext$. The shock-generated entropy errors discussed in the previous section prevent the application of this method with the energy-based scheme;  therefore, we only present the results for the entropy-conserving scheme.

Figure~\ref{fig:shocktube-injection} shows the results of the test where 20\% of the energy dissipated by the shock is converted into the nonthermal component, which is tracked separately and is shown with green lines. Different columns show the results for different values of the adiabatic index of the nonthermal component, namely, $\gext = 4/3$ and 5/3. As this figure shows, for $\gext = 4/3$ the method underproduces nonthermal pressure and entropy in the post-shock region by a small amount, while for $\gext = 5/3$ the method works remarkably well.

Qualitatively, this difference can be attributed to the effect that the injection of $\eext$ changes the compressibility of the shock and to the fact that a shock is resolved by several cells, each dissipating shock energy. When $\gext$ is smaller (larger) than $\gth=5/3$, converting a fraction of dissipated energy to $\eext$ results in a lower (higher) effective adiabatic index of gas in the post-shock region, making the shock more (less) compressible. The total change in the injected $\eext$ accumulates across the region over which the shock is numerically smeared and thus the effective adiabatic index also changes cell-by-cell gradually changing gas compressibility. As the left column in Figure~\ref{fig:shocktube-injection} shows, the total change in $\eext$ does not add up correctly and results in smaller than expected injected $\eext$ for $\eext = 4/3 < \gth$. We also checked that $\gext > \gth$ leads to an overestimate of the injected $\eext$.

In contrast, when $\gext = \gth$, the effective adiabatic index stays constant and therefore the injection of $\gext$ does not affect the compressibility of the gas. As the right column in the figure shows, our injection scheme works well in this case.

\subsection{Zel'dovich pancake test}

\begin{figure*}
\centering
\includegraphics[width=\columnwidth]{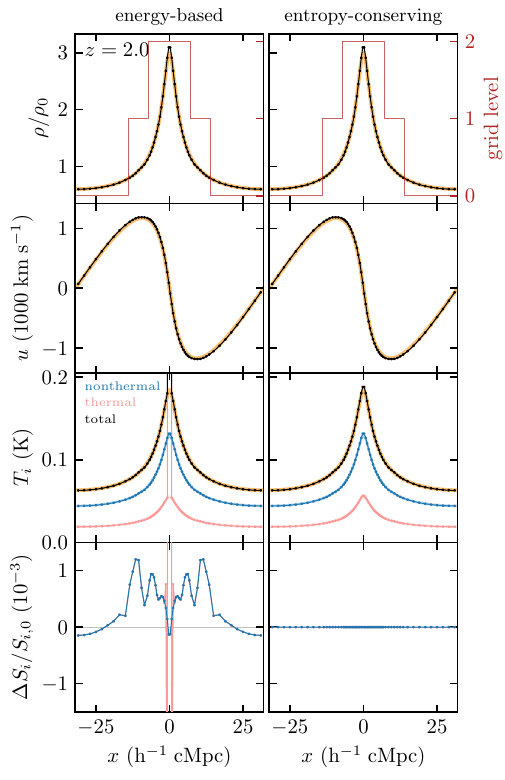}\hspace{25pt}%
\includegraphics[width=\columnwidth]{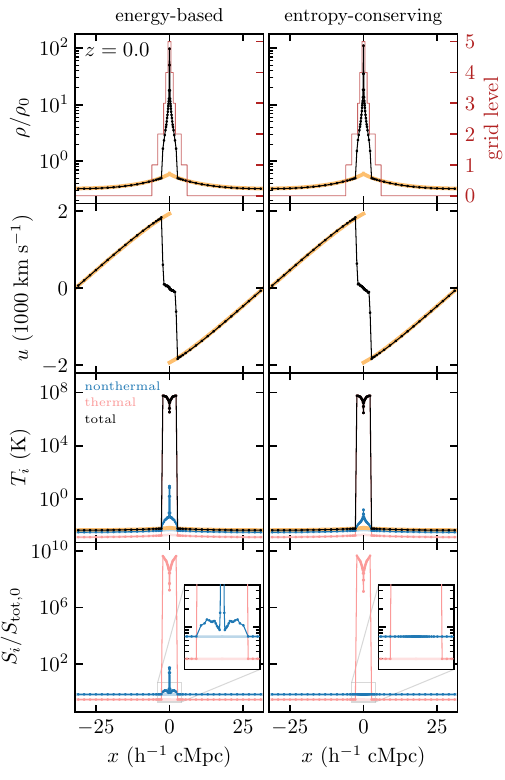}
\caption{\label{fig:zeldovich} Hydrodynamic Zel'dovich pancake test with $70\%$ of internal energy in the nonthermal form with $\gext=\gth=5/3$. The results are shown before and after the wave crossing (left and right sets of panels, respectively) and for the energy-based and entropy-conserving methods (left and right columns in each set). The connected points show simulation results, while the thick lines show the analytic solution: orange lines show gas density, velocity, and total temperature, while red and blue lines show thermal and nonthermal components. Note that the analytic solution is valid only outside the shocked regions. Thin red lines in the top panels show the grid refinement level, with the 0-th level corresponding to 32 cells per box size. The bottom panels in the left set show the relative entropy error, while in the right set, they show the absolute value of entropy normalized to the initial total entropy. The energy-based scheme conserves nonthermal entropy relatively well during the adiabatic stage (with a relative error of $<1\%$). After the wave crossing, however, the nonthermal entropy suffers from strong numerical heating at the shocks. The entropy-conserving scheme, in contrast, conserves nonthermal entropy throughout the duration of the run, with the nonthermal ``temperature'' increasing only due to adiabatic compression. }
\end{figure*}

The formation of a Zel'dovich pancake is a stringent test for an implementation of thermal and nonthermal energies because it involves both a purely adiabatic stage with extremely supersonic gas motion and the formation of strong shocks after the crossing time.

The solution for the density, $\rho(x)$, and velocity, $u(x)$, of a sine perturbation evolving from an initial redshift $z_i$ to some later redshift $z$ can be expressed in a parametric form \citep{Zeldovich.1970}:
\begin{align}
    x(q,z) &= q - \frac{1+z_{\rm c}}{1+z} \frac{\sin(kq)}{k}, \\
    \rho(q,z) &= \rho_0 \left[1 - \frac{1+z_{\rm c}}{1+z} \cos(kq) \right]^{-1}, \\
    u(q,z) &= - H_0 \frac{1+z_{\rm c}}{\sqrt{1+z}} \frac{\sin(kq)}{k},
\end{align}
where $q$ is the Lagrangian coordinate, $\rho_0$ is the average density, $k=2\pi/\lambda$ is the wavenumber of the initial perturbation, and the amplitude of the wave is described by the redshift of wave crossing, $z_{\rm c}$. 

For ease of comparisons, we choose parameters similar to \citet{Bryan.etal.1995}, \citet{Trac.Pen.2004}, and \citet{Springel.2010}: $\lambda = 64 h^{-1}\Mpc$ and $z_{\rm c}=1$, and initialize the test at $z_{\rm i}=100$. We add a nonthermal energy component with $\gext=5/3$ and initialize $\eth$ and $\eext$ so that their initial entropies are constant with values corresponding to the average temperatures of $T_{\rm i} = (\mu m_{\rm p}/k)S_{\rm i}\rho_0^{\gamma-1}=30\K$ and $70\K$ for thermal and nonthermal components, respectively (where $\mu m_{\rm p}$ is the average particle mass in units of the proton mass). 

Both entropy components are expected to be conserved until the formation of the shocks at the wave crossing, and the solution for the temperatures is therefore
\begin{equation}
    T(q,z) = T_{\rm i} \left[ \left(\frac{1+z}{1+z_{\rm i}}\right)^3 \frac{\rho(q,z)}{\rho_0} \right]^{\gamma-1}.
\end{equation}

Figure~\ref{fig:zeldovich} shows the results of the Zel'dovich test  before ($z=2$) and after ($z=0$) the formation of shocks. Both energy-based and entropy-conserving schemes produce good results for the gas density and velocity: during the adiabatic stage, $\eth$ and $\eext$ are negligible and thus their treatment does not affect the solution, while the agreement after the wave crossing indicates that both methods capture the formation of physical shocks at $z=z_{\rm c}=1$. 

In contrast, the solutions for the thermal and nonthermal components are quite different. 
As the lower left panel shows, before the crossing redshift, the energy-based scheme conserves nonthermal entropy to subpercent level, while the thermal entropy suffers from strong numerical heating at the very center of the wave. The entropy-conserving scheme, in contrast, ensures entropy conservation for both components until the shock forms. After shock formation, both schemes produce similar results for the thermal entropy, while the solution for the nonthermal entropy in the energy-based scheme suffers from spurious heating at the shocks, resulting in an increase in the nonthermal entropy by a factor of $\sim$2 up to $\sim$50 at the very center of the caustic. The error has the same origin as the errors discussed in Section~\ref{sec:tests:shocktube} and is completely absent in the entropy-conserving scheme.

The differences in thermal energy evolution before crossing illustrate the effects of the criteria selected to synchronize $\eth$ and $\eext$ with $\etot-\ekin$. Large value of thermal energy in the center in the energy-based scheme occurs because thermal energy is reset there from $\etot$, while $\eth$ is still smaller than $\etot$ by orders of magnitude. The truncation error of the scheme (typically $\sim 10^{-3}\text{--}10^{-2}\;\etot$) thus  propagates into $\eth$ as it is reset via $\eth=\etot-\ekin-\eext$, leading to orders of magnitude increase in the temperature at the wave center.
This resetting in the center occurs because the criterion $(\etot-\ekin)/\etot > \eta$ can be satisfied near the center of the wave, where $u \sim 0$ even though the flow is highly supersonic in the adjacent cells. Accounting for neighboring cells in the criterion (Equation~(\ref{eq:b95-sync2})) or setting the threshold $\eta$ to a larger value does not prevent this issue, but only delays its onset because the error in $\etot-\ekin$ accumulates with time. 

The entropy-conserving scheme adopts additional criteria based on the Mach number and velocity of the shocks in the Riemann solution (Equations~(\ref{eq:crit-Mach}) and (\ref{eq:crit-vshock})) that filter out such cases, ensuring entropy conservation until the wave crossing. In addition, given that during the adiabatic stage the conservation of entropy is enforced instead of $\etot$, the errors in $\etot-\ekin$ do not accumulate.

\section{Isolated galaxy simulations}
\label{sec:galaxy}

To test different methods for modeling nonthermal energies in a more realistic environment, we use simulations of an isolated $\sim$\Lstar~galaxy. Specifically, for the tests presented below, we use a snapshot from our fiducial simulation explored in \citet{Semenov.etal.2017,Semenov.etal.2018,Semenov.etal.2019} as the initial conditions. Below, we briefly describe the aspects of this simulation that are most relevant to the current study and refer the reader to our previous papers for more details.

One of the key ingredients of these simulations is the dynamic model for unresolved turbulence, which is implemented as a nonthermal energy. Our implementation \citep{Semenov.etal.2016} is based on the ``shear-improved'' model of \citet{Schmidt.etal.2014}. In this model, the turbulent energy on unresolved scales, $\eturb$, is modeled as a nonthermal energy component with an adiabatic index of $\gamma = 5/3$, with source and sink terms that describe the turbulent cascade from the resolved velocity fluctuations and dissipation of turbulence on the local cell-crossing time. Apart from these source and sink terms, the subgrid turbulent energy is equivalent to $\eext$ as introduced and tested in Sections~\ref{sec:schemes} and \ref{sec:tests}.

\begin{figure*}
\centering
\includegraphics[width=0.8\textwidth]{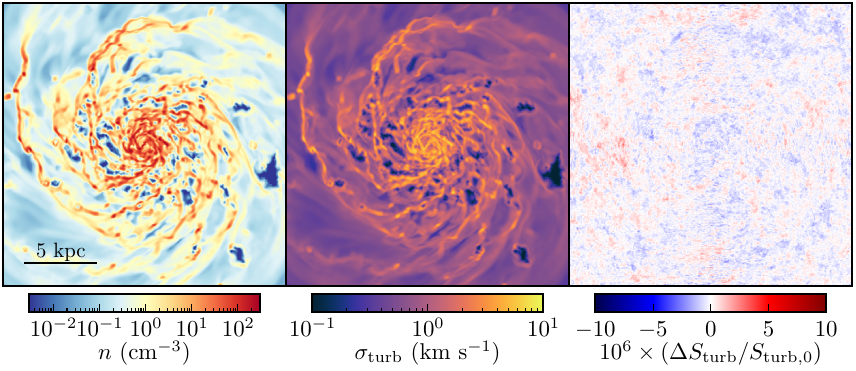}
\caption{\label{fig:maps-constS} Gas density (left), velocity dispersion of subgrid turbulence (center) and the error of the subgrid turbulence entropy (right) in an isolated disk galaxy simulation after $\sim300\Myr$ of evolution. The subgrid turbulence is modeled using the entropy-conserved scheme. In this test, subgrid turbulence is initialized such that its entropy is initially constant, and then the galaxy simulation is evolved with all the source and sink terms of the subgrid turbulence model turned off, except for the $PdV$ term. The right panel illustrates the long-term conservation of the entropy of subgrid turbulence, which remains constant to machine precision ($\sim10^{-6}$ for a single-precision floating point variable) after $\sim300\Myr$ of evolution. }
\end{figure*}

The initial conditions for our isolated galaxy simulations are taken from the AGORA code comparison project \citep{Kim.etal.2014}. The AMR grid cells are adaptively refined when the gas mass in a cell exceeds $\sim 8300\Msun$ until the minimal cell size of $\Delta = 40\pc$ is reached. Radiative cooling and heating of gas are modeled using the \citet{Gnedin.Hollon.2012} model, assuming constant metallicity at the solar value and a constant radiation background with the H$_2$ photodissociation rate in the Lyman--Werner bands of $10^{-10}\;{\rm s^{-1}}$ \citep{Stecher.Williams.1967}. We also adopt a model for dense gas shielding, which is calibrated against radiative transfer simulations of the ISM \citep[the ``L1a'' model from][]{Safranek-Shrader.etal.2017}. 

Local star formation is parametrized via the star formation efficiency per freefall time, $\rhoSFR = \epsff \rho/\tff$, with a fixed value of $\epsff=1\%$. Star-forming gas is defined by using a threshold either in density (Section~\ref{sec:galaxy:adiabatic}) or in the local virial parameter (Section~\ref{sec:galaxy:full}), where the latter is computed from the local turbulent velocity, $\sturb = \sqrt{2 \eturb / \rho}$, by using the definition from \citet{Bertoldi.McKee.1992}:
\begin{equation} 
\label{eq:avir}
    \avir \equiv \frac{5 \stot^2 R}{3GM} \approx 9.35 \frac{ (\stot/10\kms)^2 }{ (n/100\cc) (\Delta/40 \pc)^2}.
\end{equation}

Stellar feedback is implemented by injecting radial momentum and thermal energy in the amounts calibrated against high-resolution simulations of SN remnants evolution in a nonuniform medium \citep{Martizzi.etal.2015}, with the momentum boosted by a factor of 5 to account for the effects of SN clustering \citep[e.g.,][]{Gentry.etal.2017,Gentry.etal.2018} and cosmic rays \citep{Diesing.Caprioli.2018}. To approximate the effects of pre-SN feedback, we start the injection of momentum from the moment of star particle formation and continue the injection at a constant rate for 40 Myr. The injection rate is set by the total number of SNe occurring for a given star particle, which we compute assuming the \citet{Chabrier.2003} IMF. We also account for stellar mass loss by using the \citet{Leitner.Kravtsov.2011} model.

\subsection{Adiabatic subgrid turbulence energy}
\label{sec:galaxy:adiabatic}

We start by demonstrating that, in a realistic ISM with a wide variation in density and temperature, our implementation of the entropy-conserving scheme ensures a long-term conservation of nonthermal entropy. To this end, we initialize the subgrid turbulent energy such that its entropy is constant and turn off all its source and sink terms. As such a model for turbulence is not realistic, we use a star formation threshold not in $\avir$ but in density, defining all gas with $n>\nsf=100\cc$ as star-forming. We select this threshold value because, for our simulated galaxy, it results in roughly the same amount of star-forming gas as the fiducial $\avir$ threshold \citep{Semenov.etal.2018}. 
For this test, we also modified the original star formation and refinement algorithms such that the nonthermal entropy of gas in a cell is conserved when a star particle is formed or the cell is refined or de-refined.

\begin{figure*}
\centering
\includegraphics[width=0.8\textwidth]{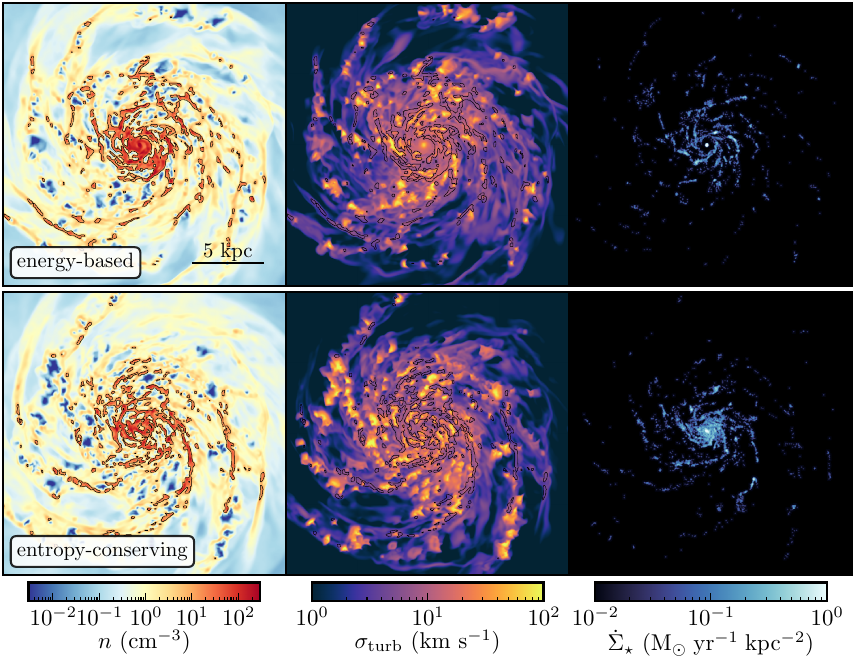}
\caption{\label{fig:maps} Results of the isolated galaxy simulation with the full subgrid turbulence model and star formation prescription coupled with the predicted turbulent velocities. The columns show the mid-plane slices of density, $n$, and subgrid turbulent velocity, $\sturb$, and the surface density of SFR, $\SSFR$, computed using particles younger than $30\Myr$. Different rows show simulations where thermal energy and subgrid turbulence are modeled by using either the energy-based or entropy-conserving scheme. To guide the eye, black contours in the $n$ and $\sturb$ maps show $n=10\cc$ density isocontours. The $\SSFR$ maps show the most apparent difference between the two schemes: the SFR in the run with the energy-based scheme is more clumpy and strongly suppressed near the galaxy center. }
\end{figure*}

Figure~\ref{fig:maps-constS} shows the maps of gas density, subgrid turbulent velocity, and entropy error after 300 Myr of evolution. In the absence of source and sink terms, the nonthermal entropy maintains its initial value to machine precision, $\sim 10^{-6}$ for the single-precision floating point used in this test. 

We do not show the results for the energy-based scheme because there the entropy errors discussed in Section~\ref{sec:tests:shocktube} quickly accumulate resulting in a steady increase in the turbulent pressure that eventually stabilizes the disk. In a simulation with a more realistic processes, such runaway heating does not occur because the cooling term in the equation for the nonthermal energy leads to efficient dissipation of erroneously generated turbulent energy. Therefore, in the next subsection, we compare the schemes with all the sink and source terms activated.

\subsection{Full subgrid turbulence model}
\label{sec:galaxy:full}

As a more realistic test, we rerun our galaxy simulation including all source and sink terms in the subgrid turbulence model. We also use a star formation threshold based on the subgrid virial parameter, $\avir<\avirsf=10$, which was our fiducial choice in \citet{Semenov.etal.2017,Semenov.etal.2018,Semenov.etal.2019}.

Figure~\ref{fig:maps} compares the maps of gas density, subgrid turbulent velocity, $\sturb$, and SFR surface density in simulations with the energy-based and entropy-conserving schemes. Although the overall distributions of $n$ and $\sturb$ are quite similar, the figure reveals interesting differences. With the entropy-conserving scheme, the overall density structure becomes more flocculent, especially near the disk center. The difference in the $\SSFR$ map is even more striking: while with the energy-based scheme the $\SSFR$ distribution is rather clumpy and has a strong deficiency in the central $\sim1\kpc$, in the simulation with the entropy-conserving scheme, $\SSFR$ is smoother and is not suppressed near the center.

\begin{figure}
\centering
\includegraphics[width=\columnwidth]{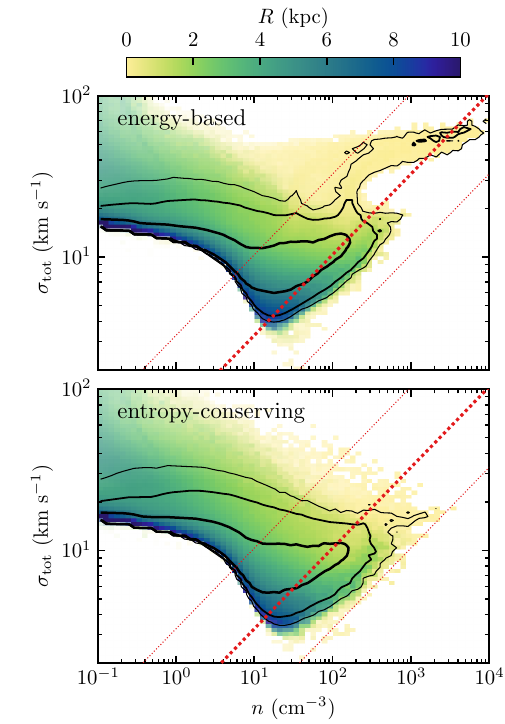}
\caption{\label{fig:n-sigma} Joint distribution of gas density and total subgrid velocity dispersion, which includes both subgrid turbulence and thermal sound speed, $\stot = \sqrt{\sturb^2+\cs^2}$. The contours show 68\%, 95\%, and 99\% of the gas mass and the color indicates the average galactocentric radius of cells within a pixel. The thick dotted line shows the star formation threshold, $\avirsf = 10$, while the thin lines show the values of $\avir=1$ and 100 for reference. To reduce noise, the distributions are averaged over 21 snapshots spaced 10 Myr apart. Switching to the entropy-conserving scheme reduces $\stot\approx\sturb$ in dense gas ($n>10\cc$) by a factor of $\sim$1.5 and completely removes the horizontal feature at $\sturb>30\kms$ and $n>100\cc$ that corresponds to the central region of the disk. }
\end{figure}

To make the comparison more quantitative and explain these differences, Figure~\ref{fig:n-sigma} shows joint distributions of gas in the \ns~plane colored according to the galactocentric radius of the cells and with the thick dotted line showing the star formation threshold. 

In the warm and diffuse part of the ISM with $n<10\cc$, the overall shapes of these distributions are similar. In most of such gas, $\stot$ is dominated by the sound speed, $\cs \sim 4\text{--}20\kms$, which sets the lower envelope of the distribution at $n<10\cc$, reminiscent of the shape of the $n$--$T$ diagram at such densities. We checked that the distributions of $\sturb$ alone are also quite similar at $n<10\cc$ in both schemes.

In contrast, there are two interesting differences in the cold and dense ISM, $n>10\cc$, where subgrid turbulence dominates over thermal energy. First, the turbulent velocities in the main part of the distribution decrease by a factor of $\sim$1.5 when the entropy-conserving scheme is used. Second, with the energy-based scheme, the gas near the disk center manifests itself as a prominent feature with $n>100\cc$ and $\sturb \sim 30\text{--}50\kms$, while with the entropy-conserving scheme, this feature disappears, and $\sturb$ in the central region becomes a continuation of the rest of the distribution.

The source of these differences is the production of $\eturb$ in radiative shocks via the mechanism described in Section~\ref{sec:tests:shocktube}. The average distribution of cold gas in the \ns~plane is set by the competition between turbulence production and its decay on the local cell-crossing time. While both simulations include physical production terms---adiabatic compression and sourcing by fluctuating velocities---the numerical heating discussed in Section~\ref{sec:tests:shocktube} also contributes to the production of $\eturb$ when the energy-based scheme is used. The fact that in most of the cold ISM the increase in $\sturb$ is only moderate indicates that this heating mechanism does not dominate over the physical production terms. The only exception is the disk center, where switching to the entropy-conserving scheme leads to a decrease in $\sturb$ by a factor of $\sim$3, which corresponds to a decrease in $\eturb$ by an order of magnitude.

Even though in the context of our subgrid turbulence model the excess of $\eturb$ in the disk center is purely numerical, it may reflect a physical enhancement of turbulence production in that region due to strong shear. Such an enhancement may not be captured by the adopted turbulence production term in the model, but it does appear as an enhancement of numerical dissipation of kinetic energy that is partially converted to $\eturb$ (see Section~\ref{sec:discussion:turbulence} for further discussion). Therefore, here we solely present the difference between the two schemes of modeling nonthermal energies and leave the detailed investigation of the behavior in the disk center to a future study.

The decrease in $\sturb$ in the cold and dense gas leads to significant changes in the SFR distribution. For example, as $\sturb$ drops near the disk center, gas can cross the star formation threshold more easily in the run with the entropy-conserving scheme in striking contrast with the strongly suppressed SFR in the central $\sim1\kpc$ in the run with the energy-based scheme (recall the right panels in Figure~\ref{fig:maps}). At larger galactocentric radii, distribution of star-forming regions also becomes different. As Figure~\ref{fig:n-sigma} shows, when the entropy-conserving scheme is used, a larger fraction of gas from the main part of the distribution (i.e., excluding the feature corresponding to the disk center) reaches low virial parameters, $\avir<\avirsf=10$, and the typical densities of such star-forming gas become lower.

\begin{figure}
\centering
\includegraphics[width=\columnwidth]{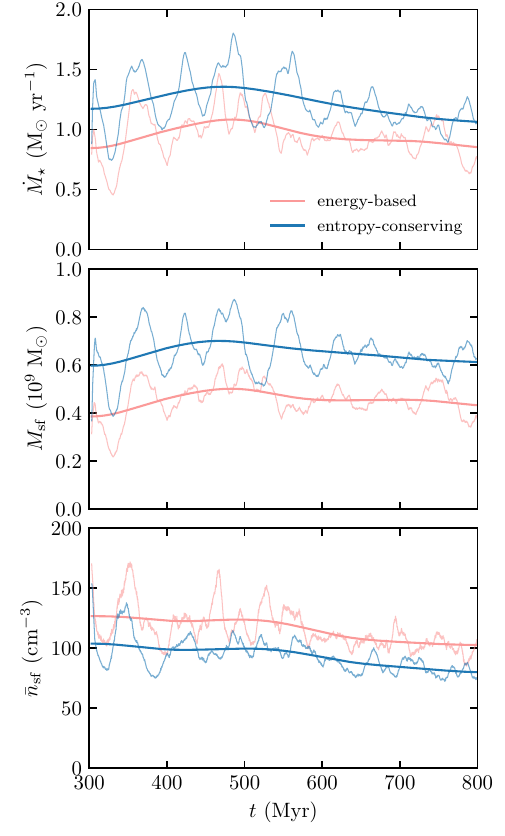}
\caption{\label{fig:sfr} Effect of thermal and nonthermal energy modeling on the star formation in the average disk, i.e., at $R>1\kpc$. The panels show the total star formation rate, mass of star-forming gas, and its average density. When the entropy-conserving scheme is used, a lower $\sturb$ in dense gas (see Figure~\ref{fig:n-sigma}) results in a larger amount of star-forming gas but also in a lower average density of such gas. These trends partially cancel out, leading to only a moderate increase in the SFR. }
\end{figure}

Figure~\ref{fig:sfr} quantifies the effects of the turbulent energy scheme on the SFR, total mass of star-forming gas, $\Msf$, and its average density, $\bar{n}_{\rm sf}$, at $R>1\kpc$. These three quantities are related via
\begin{equation}
\label{eq:sfr}
\begin{split}
\SFR &= \int \rhoSFR dV = \int_{\rm sf} \frac{\epsff \rho}{\tff} dV =\\
     &=\epsff \Msf \left\langle \frac{1}{\tff} \right\rangle_{\rm sf} \propto \epsff \Msf \bar{n}_{\rm sf}^{0.5},
\end{split}
\end{equation}
where $\epsff=1\%$ and the average density of star-forming gas is defined from $\langle 1/\tff \rangle_{\rm sf}^{-1} \equiv \sqrt{ 3\pi / (32 G \mu m_{\rm p} \bar{n}_{\rm sf} )}$, assuming $\mu=1$ and $\langle ... \rangle_{\rm sf}$ denoting mass-weighted averages over star-forming regions.

As the figure shows, the effect of the scheme on these quantities is quite significant, at the level of $\sim10\%\text{--}30\%$. In the entropy-conserving scheme, $\nsf$ is smaller and $\Msf$ is larger, which is consistent with the change in the PDF in Figure~\ref{fig:n-sigma}. These changes partially cancel in the total SFR, as the larger mass of star-forming gas is partly offset by the lower density and longer freefall time of star-forming regions. The total SFR in the run with the entropy-conserving scheme thus increases only moderately. 

The direction of these trends is consistent with the model for the origin of global gas depletion times, $\tglob \equiv \Mg/\SFR$, and SFR from continuous and rapid gas cycling between actively star-forming and diffuse, non-star-forming states in the ISM \citep{Semenov.etal.2017}. In this model, the global depletion time is the sum of the times that gas spends in each of these states, which can be written as $\tglob = \Nc \tnsf + \tau_{\rm ff}/\epsff$, where $\Nc$ is the total number of cycles required for gas depletion, $\tnsf$ is the average residence time of gas in the non-star-forming state, and $\tau_{\rm ff}/\epsff$ is the average depletion time of the star-forming gas. Switching to the entropy-conserving scheme reduces the production of $\eturb$, making it easier for gas to cross the $\avir\propto\sturb^2/n$ threshold, which leads to shorter $\tnsf$ and longer $\tau_{\rm ff}$. These trends counteract in the expression for $\tglob$, but they add up in the expression for the star-forming gas mass fraction: $\fsf = (\tau_{\rm ff}/\epsff)/\tglob = (\epsff \Nc \tnsf/\tau_{\rm ff} + 1)^{-1}$. As a result, the increase in $\SFR = \Mg/\tglob$ is smaller than the increase in $\Msf = \fsf \Mg$ as $\Mg$ stays roughly constant over the considered time interval.

\section{Discussion}
\label{sec:discussion}

\subsection{Implications for galaxy formation simulations}

The results presented in the previous sections show that the choice of the method to model thermal and nonthermal energy components matters. In particular, the energy-based scheme can lead to a spurious generation of both thermal and nonthermal energy components. This was apparent in the Zel'dovich pancake collapse test, which showed significant spurious temperature increase at the center of the pancake right before crossing. Similarly, \citet{Villasenor.etal.2020} showed that the mean IGM temperature in their simulations is sensitive to how thermal energy of the gas is modeled and to the parameters of the dual-energy scheme (see their Section 3.4 and Figures 5 and 6).

This difference arose because shock heating of gas in collapsed regions of virialized halos could be reduced or suppressed by the condition for synchronizing thermal energy followed independently with the value computed from the total energy. The results  of \citet{Villasenor.etal.2020} indicate that these issues can be corrected by the proper choice of thermal energy synchronization parameters, meaning that both energy- and entropy-based schemes can accurately and robustly model thermal energy. However, our results show that the choice of such scheme for any \emph{nonthermal} component(s), such as subgrid turbulence or cosmic rays,  does make a difference. 
This is because nonthermal components are necessarily evolved separately from the total energy in the entire simulation domain, while such treatment of the thermal energy alone is only needed in a limited volume corresponding to highly supersonic flows. 

Many modern fluid dynamics codes employ different schemes to separately model the coherent bulk motion of the fluid and small-scale motions \citep{Trac.Pen.2004,Springel.2010,Duffell.MacFadyen.2011,Duffell.MacFadyen.2015,Hopkins.2015,Duffell.2016}. Such ``moving-mesh'' approaches alleviate the inaccuracy of the thermal energy calculation from the total energy by absorbing a significant fraction of kinetic energy into the motion of the mesh.  However, our results are equally relevant for these approaches because they indicate that the entropy-conserving scheme is the preferred choice for modeling the energies of nonthermal components. Indeed, for the energy-based formulation, the errors in the nonthermal entropy originate from the nonadiabatic part of the explicit $PdV$ source term inside numerically smeared shocks which occur both in Eulerian and moving-mesh fluid dynamics schemes. 

How much the choice of the scheme matters depends on the problem at hand. In the relatively dense parts of the ISM, where cooling is efficient, the spurious generation of energy near shocks may be inconsequential as it is dissipated efficiently. On the other hand, as we showed in the previous section, this choice may result in nonnegligible changes in the star formation rate of a galaxy, the structure of the ISM, and in particular qualitatively different star formation in the central regions of the simulated galaxy. 

The origin of the entropy error is closely related to the closure problem of the shock solution in the presence of nonthermal energy components \citep{Gupta.etal.2021}. The classical Rankine--Hugoniot jump conditions become insufficient and require additional relations to define the behavior of each of the extra nonthermal components. In the methods explored in our paper, these closure relations are effectively set by the choice of the method to evolve different components and the algorithm that synchronizes these components with the total energy (e.g., Equation~(\ref{eq:eth-sync}) or Section~\ref{sec:schemes:injection}).

The results of \citet{Gupta.etal.2021} and our findings indicate that the energy-based implementation implicitly defines a closure relation that deviates from an adiabatic behavior across shocks despite the fact that the underlying equation describes adiabatic evolution. More importantly, the behavior of nonthermal components becomes dependent on the properties of the shock and on the details of the numerical implementation. This is because entropy errors originate from the nonadiabatic excess of the $PdV$ source term in the region over which the shock is numerically smeared. The structure of the smeared shock depends on both physical properties of the shock and details of the numerical method, and so will the entropy error in the post-shock region. This error thus can be avoided by either getting rid of the source term altogether, as we do in this study \citep[see also][]{Kudoh.Hanawa.2016,Gupta.etal.2021}, or by imposing an explicit closure relation at shocks, as suggested by \citet{Gupta.etal.2021}. 

Although the entropy-conserving scheme is a well-posed solution to the above problem, \citet{Kudoh.Hanawa.2016} and \citet{Gupta.etal.2021} reported issues arising when advecting a ``pressure-balance’’ mode, where gas density, velocity, and total pressure are all constant, while the fractions of thermal and nonthermal pressure components experience a jump. We have repeated such a test and found that our implementation of the entropy-conserving scheme performs extremely well in this test: density, velocity, and total pressure are all preserved to machine precision (see Appendix~\ref{app:pressure-balance}). Thus, the performance in the ``pressure-balance’’ test likely depends on the details of scheme implementation.

Regardless of numerical issues, a strong argument for the entropy-based scheme for nonthermal energy components can be made in models that include  the injection or dissipation of energy in shocks. 
For example, both turbulence and cosmic ray energy can be generated at shock fronts and, as we discussed in Section~\ref{sec:schemes:injection}, such generation can be handled with ease in the entropy-conserving scheme.

This approach can be especially advantageous for sourcing nonthermal energies on radiative shocks. The existing methods based on on-the-fly shock-finding algorithms estimate dissipated energy by measuring the pressure jump across the shock. This can be inaccurate when the pressure in the post-shock region is subject to significant radiative losses. This issue can be alleviated in the proposed method because the energy dissipated by shocks at all cell interfaces can be converted to nonthermal components before radiative losses occur. We leave a more detailed investigation of this issue and a comparison of different methods to future study. 

Finally, the entropy-conserving scheme can provide another significant advantage for modeling subgrid turbulence because the resolved turbulent energy dissipated by numerical viscosity can be captured into the subgrid turbulent energy {\it implicitly}, as we outline in the next section. This allows one to avoid explicit modeling via source terms that require calibration and can be rather uncertain.

\subsection{Sourcing subgrid turbulence via capturing kinetic energy dissipated by numerical viscosity}
\label{sec:discussion:turbulence}

Explicit modeling of unresolved turbulence in galaxy formation simulations can significantly improve the modeling of processes that depend on the small-scale turbulent structure of the ISM, such as star formation, mixing and transport of metals, and the effect of gas clumping on cooling and chemical reaction rates. Such subgrid models of turbulence require a prescription for the energy transfer from the resolved flow to unresolved scales.

For example, in the Large Eddy Simulation (LES) approach adopted in Section~\ref{sec:galaxy:full} (see \citealt{Garnier.etal} and \citealt{Sagaut} for reviews), such source terms are modeled using explicit closure relations, which can be calibrated against direct high-resolution simulations of turbulence \citep[e.g.,][]{Schmidt.Federrath.2011}. These closures are the major source of uncertainty in modeling subgrid turbulence, especially when the onset of the turbulent cascade is not sufficiently resolved or the local flow is significantly different from the direct simulations used to calibrate the closures.

On the other hand, the strict enforcement of entropy conservation opens up a new, more generic way for modeling the energy cascade to unresolved scales via tracking the local dissipation of the resolved kinetic energy. Indeed, as our results show, this scheme can be used to accurately follow adiabatic evolution of thermal and nonthermal energies during each hydrodynamics step. In regions where thermal energy can be computed reliably via the difference of the total and kinetic energies, $\etot-\ekin$, the change in this energy during a single step can be compared to the adiabatically evolving sum $\eth+\eturb$ modeled separately using the entropy-conserving scheme.\footnote{Note that in the presence of cooling, the dissipation of energy can be accounted for in such a comparison during a single step.} The difference between the two estimates will then correspond to the total kinetic energy dissipated by numerical viscosity in each cell during the step. 

In most galaxy simulations, this dissipated energy is converted directly into heat and then quickly radiates away. In the real ISM, however, the energy is transferred to turbulent motions on small scales and decays on a longer timescale, close to the local eddy-turnover time. The above method can be used to capture this energy and to convert it into heat on a physically motivated timescale, e.g., turbulent cell-crossing time with an order-of-unity pre-factor that can be calibrated in direct simulations of turbulence as in LES simulations. 

Even though the dissipation of kinetic energy in this method is purely numerical and happens on the scale of a few computational cells, empirical evidence suggests that the \emph{rate} of dissipation is correct. Indeed, turbulent box simulations carried out with different codes generally find turbulent spectra consistent with theoretical expectations, suggesting that the rate of the energy transfer on resolved scales is modeled correctly \citep[e.g.,][]{Kritsuk.etal.2011}. Given that in developed turbulence, this energy transfer rate is equal to the dissipation rate on the viscous scale, the rate of numerical dissipation must also be equal to the physical dissipation rate. 

One additional issue that needs to be considered in such an approach is that not all kinetic energy dissipated by numerical viscosity should source the subgrid turbulence. For example, numerical viscosity can dissipate energy in laminar shearing flows, but no turbulence should be generated in this case. In the context of the LES approach, this motivates the usage of the ``shear-improved'' versions of the schemes, in which the bulk flow of the gas is factored out from the source terms of subgrid turbulence \citep{Leveque.etal.2007,Schmidt.etal.2014}.
Conversely, a fraction of the energy dissipated by resolved shocks can drive turbulent motions in the post-shock region instead of being fully thermalized. 
These examples show that an approach described above should include a model for how the dissipated energy is apportioned between thermal and subgrid turbulence energy based on the local flow properties, such as the shear-improved approach and the scheme outlined in Section~\ref{sec:schemes:injection}.

The idea of unresolved turbulence sourcing by numerical dissipation is a part of a more general assumption upon which the concept of the Implicit Large Eddy Simulations (ILES) is based (see, e.g., \citealt{ILES} for a review): as long as a hydrodynamics solver satisfies the basic physical properties of the equations of hydrodynamics, such as conservation of mass, momentum, and energy, as well as causality and positivity, the effect of truncation errors at the resolution scale is equivalent to the net effect of unresolved scales. In this context, the approach outlined above captures the energy transfer to unresolved scales for the explicitly modeled subgrid turbulence without explicit and uncertain terms and closure relations. Instead, the energy transfer  is implicitly handled by the hydrodynamic solver. The subgrid turbulent energy can then be used to model star formation, metal diffusion, and gas clumping factor in exactly the same way as in the LES simulations.

\section{Summary and Conclusions}
\label{sec:summary}

We have investigated the accuracy of energy-based and entropy-conservative schemes in modeling nonthermal energy components. These schemes were initially introduced for modeling thermal energies of highly supersonic flows where $\eth$ is strongly subdominant to the total and kinetic energies and therefore cannot be accurately computed as the difference between the two. The energy-based scheme numerically solves the explicit equation for the energy density evolution, which cannot be written in conservative form. The entropy-conserving scheme, on the other hand, uses a conservative equation for modified entropy and can thus conserve the entropy of an energy component to machine precision.

We have examined the performance of the schemes in following the subgrid turbulence and cosmic ray energy in the standard shock tube and Zel'dovich pancake tests and simulations of a realistic isolated \Lstar~galaxy. 
Our results can be summarized as follows. 

\begin{enumerate}
    \item The energy-based scheme results in a spurious generation of nonthermal energy on shocks in the shock tube and  Zel'dovich pancake tests, while the entropy-conserving method evolves the energy adiabatically to machine precision (see Figures~\ref{fig:shocktube}, \ref{fig:shocktube-compressive}, and \ref{fig:zeldovich}). 
    
    \item The magnitude of the nonthermal entropy error in the post-shock region increases with the adiabatic index of the nonthermal component and with the compression ratio of the shock (see Figure~\ref{fig:S-error}). The latter is particularly relevant for the cold ISM where shocks can be strongly compressible due to efficient radiative cooling.
    
    \item In simulations of an isolated \Lstar~galaxy with a turbulence-based star formation prescription, switching between the energy-based and entropy-conserved schemes results in a qualitative change in morphology, in particular, a large qualitative difference in star formation in the center of the galaxy (Figure~\ref{fig:maps}), and $\approx 20\text{--}30\%$ change in the star formation rate away from the center (Figure~\ref{fig:sfr}). 
    
    \item We outline and test a simple, physical method for the injection of nonthermal energy on shocks. This method can be used in conjunction with the entropy-conserving scheme and can simplify the implementation of, e.g., injection of cosmic rays and driving of subgrid turbulence by shocks (see Section~\ref{sec:schemes:injection} and Figure~\ref{fig:shocktube-injection}).  
\end{enumerate}

Finally, we discuss how the entropy-conserving scheme can provide a straightforward way to capture the kinetic energy dissipated by numerical viscosity into the subgrid turbulent energy {\it implicitly}, without explicit source terms that require calibration and can be rather uncertain. It will be interesting to investigate the performance of such an approach using idealized simulations of turbulence and realistic galaxy formation simulations.

Although both methods have been shown to perform well in modeling the evolution of thermal energy, our results indicate that the entropy-conserving scheme is a preferred choice for modeling nonthermal energy components. This conclusion is equally relevant for Eulerian and moving-mesh fluid dynamics codes.

\acknowledgments
We would like to thank Nick Gnedin, Peng Oh, Tsun Hin Navin Tsung, and Siddhartha Gupta for their insightful comments.
We are also grateful to the anonymous referee whose comments helped to improve this paper.
Support for V.S. was provided by NASA through the NASA Hubble Fellowship grant HST-HF2-51445.001-A awarded by the Space Telescope Science Institute, which is operated by the Association of Universities for Research in Astronomy, Inc., for NASA, under contract NAS5-26555. A.K. was supported by the National Science Foundation grants AST-1714658 and AST-1911111 and NASA ATP grant 80NSSC20K0512.
This work was completed in part with resources provided by the University of Chicago Research Computing Center and by the NASA High-End Computing (HEC) Program through the NASA Advanced Supercomputing (NAS) Division at Ames Research Center.
The analyses presented in this paper were greatly aided by the following free software packages: {\tt NumPy} \citep{numpy_ndarray}, {\tt SciPy} \citep{scipy}, {\tt Matplotlib} \citep{matplotlib},  and {\tt yt} \citep[][]{yt}. We have also used the Astrophysics Data Service (\href{http://adsabs.harvard.edu/abstract_service.html}{\tt ADS}) and \href{https://arxiv.org}{\tt arXiv} preprint repository extensively during this project and the writing of the paper.

\appendix

\section{Advection of a pressure-balance mode}
\label{app:pressure-balance}

\begin{figure*}
\centering
\includegraphics[width=\textwidth]{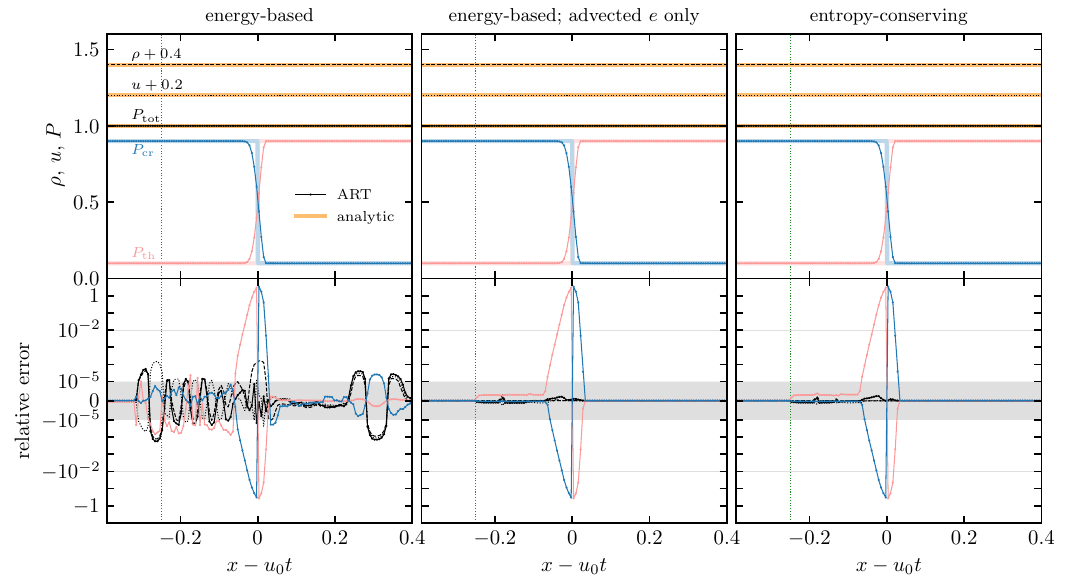}
\caption{\label{fig:pressure-balance} Results of the ``pressure-balance'' test at $t=0.25$. Different columns correspond to different schemes, from left to right: the energy-based scheme described in Section~\ref{sec:schemes}, the energy-based scheme that does not synchronize with the solution for $\etot$, and the entropy-based scheme. The initial conditions are the same as in \citet{Gupta.etal.2021}: $(\rho,u,\Pth,P_{\rm cr}) = (1,1,0.1,0.9)$ and $(1,1,0.9,0.1)$ for the left and right initial states, respectively, and the initial location of the discontinuity is marked by the vertical green dotted line. The adiabatic indices of thermal and cosmic ray components are 5/3 and 4/3, respectively. To avoid the overlap in the top panels, the values of $\rho$ and $u$ are shifted by 0.4 and 0.2, respectively. The shaded gray area in the bottom panels shows the range where the scale of the $y$-axis is linear; the horizontal gray lines indicate the relative error of $1\%$. }
\end{figure*}

\citet{Kudoh.Hanawa.2016} and \citet{Gupta.etal.2021} compared the performance of nonconservative and conservative schemes for modeling cosmic rays and reported issues of the latter scheme in the ``pressure-balance’’ test. In this test, density, velocity, and total (thermal plus nonthermal) pressure are constant throughout, but the fractions of pressure contributed by the thermal and nonthermal components experience a jump at $x=0$. Given that pressure and velocity are constant, no force should be generated and the jump in the pressure fractions should simply be advected, while other quantities should remain constant. However, \citet{Kudoh.Hanawa.2016} and \citet{Gupta.etal.2021} found that their implementation of the conservative scheme produces numerical artifacts in this test, while \citet{Gupta.etal.2021} also found that their nonconservative energy-based method performs better. 

The results of this test with our implementation of the energy-based and entropy-conserving schemes are shown in Figure~\ref{fig:pressure-balance}. 
Both our schemes perform extremely well in this test: gas density, velocity, and total pressure are all conserved to high precision, even though the jump in the pressure components becomes smeared over several cells. In particular, our entropy-conserving scheme can model advection of the pressure-balance mode to machine precision, in contrast to the results reported by \citet{Kudoh.Hanawa.2016} and \citet{Gupta.etal.2021}.

Why does our entropy scheme perform in this test better than similar models from previous literature? Not surprisingly, the results of this test strongly depend on the modeling of energy or entropy advection. Indeed, the $PdV$ source term should stay exactly 0 as long as gas velocity stays constant, and therefore any spurious pressure gradients can originate only from the inconsistencies in the advection fluxes of different energy components. These inconsistencies can arise when different components are modeled using different schemes. For example, in the entropy-based methods discussed by \citet{Kudoh.Hanawa.2016} and \citet{Gupta.etal.2021}, only the nonthermal component is modeled using an entropy-conserving scheme, while the thermal component is computed from the solution for the total energy. In contrast, in our implementation of the entropy scheme, both entropy components are modeled consistently, using the same advection scheme, which leads to a significantly better performance in this test.

In the energy-based scheme, significant artifacts may also appear when the advection fluxes of $\eth$ and $\eext$ are modeled inconsistently with the Riemann fluxes of mass and total energy. For example, we find that not applying the $\etot$ flux correction described at the end of Section~\ref{sec:schemes:energy} produces artifacts in the gas density, velocity, and total pressure at a $>$10\% level. In contrast, this correction reduces these errors to $\lesssim 10^{-4}$ (see the left panel in Figure~\ref{fig:pressure-balance}). This error is somewhat larger than in the entropy-conserving scheme because $\eth$ is reset from the difference $\etot-\ekin-\eext$. To demonstrate this, the middle panel shows the results of the energy-based scheme without this synchronization, which only uses advected $\eth$.\footnote{This can be done by setting $\eta_1$ in Equation~(\ref{eq:b95-sync1}) to an arbitrarily large number.} In this case, the error has the same small magnitude as in the entropy-conserving scheme.

All in all, these results show that the performance of the schemes in the ``pressure-balance’’ test depends on the implementation of the advection for different energy components. In contrast to the findings of previous studies, our implementation of the entropy-conserving scheme performs well in this test.

\bibliographystyle{aasjournal}
\bibliography{}

\end{document}